\documentclass[conference]{IEEEtran}
\usepackage{cite,graphicx,amsmath,amssymb}
\usepackage{subfigure}
\usepackage{citesort}
\usepackage{fancyhdr}
\usepackage{mdwmath}
\usepackage{mdwtab}
\usepackage{balance}
\usepackage{xcolor}
\usepackage{bm}
\usepackage{xspace}
\usepackage{algorithm}
\usepackage{algorithmic}
\usepackage{multirow}
\usepackage{flafter}
\usepackage{amsthm}
\usepackage[acronym,nogroupskip,nonumberlist,nopostdot]{glossaries}
\makeglossaries
\newacronym{snr}{SNR}{signal-to-noise ratio}
\newacronym{6g}{6G}{sixth-generation}
\newacronym{em}{EM}{electromagnetic}
\newacronym{5g}{5G}{fifth-generation}
\newacronym{sre}{SRE}{smart radio environment}
\newacronym{stars}{STAR-RISs}{simultaneously transmitting and reflecting reconfigurable intelligent surfaces}
\newacronym{star}{STAR-RIS}{simultaneously transmitting and reflecting reconfigurable intelligent surface}
\newacronym{mimo}{MIMO}{multiple-input multiple-multiple}
\newacronym{cscg}{CSCG}{circularly symmetric complex Gaussian}
\newacronym{los}{LoS}{Line-of-Sight}

\DeclareMathOperator{\diag}{diag}
\DeclareMathOperator{\Diag}{Diag}
\DeclareMathOperator{\rank}{Rank}
\DeclareMathOperator{\tr}{Tr}
\newcommand{\ie}{\text{i}.\text{e}.}
\newcommand{\eg}{\text{e}.\text{g}.}

\newcommand{\sixg}{\gls{6g}\xspace}
\newcommand{\emm}{\gls{em}\xspace}

\newcommand{\sre}{\gls{sre}\xspace}
\newcommand{\starris}{\gls{star}\xspace}
\newcommand{\starriss}{\gls{stars}\xspace}

\newcommand{\los}{\gls{los}\xspace}

\newtheorem{theorem}{Theorem}

\newtheorem{lemma}{Lemma}

\newtheorem{corollary}{Corollary}

\usepackage[
top    = 1.70cm,
bottom = 1.05in,
left   = 0.63 in,
right  = 0.63 in]{geometry}

\begin{document}

\title{Joint Beamforming for STAR-RIS in Near-Field Communications}

\author{
\IEEEauthorblockN{ Li~Haochen\IEEEauthorrefmark{2}\IEEEauthorrefmark{4}, Yuanwei~Liu\IEEEauthorrefmark{3}, Xidong~Mu\IEEEauthorrefmark{3}, Yue~Chen\IEEEauthorrefmark{3}, and Pan~Zhiwen\IEEEauthorrefmark{2}\IEEEauthorrefmark{4}\IEEEauthorrefmark{1} } \IEEEauthorblockA{
\IEEEauthorrefmark{2} National Mobile Communications Research Laboratory, Southeast University, Nanjing, China\\
\IEEEauthorrefmark{3} Queen Mary University of London, London, UK\\
\IEEEauthorrefmark{4} Purple Mountain Laboratories, Nanjing, China\\
\IEEEauthorrefmark{1} Corresponding author\\
 } }
\maketitle

\begin{abstract}
A \starris aided near-field multiple-input multiple-output (MIMO) communication framework is proposed. A weighted sum rate maximization problem for the joint optimization of the active beamforming at the base station (BS) and the transmission/reflection-coefficients (TRCs) at the STAR-RIS is formulated. The resulting non-convex problem is solved by the developed block coordinate descent (BCD)-based algorithm. Numerical results illustrate that the near-field beamforming for the STAR-RIS aided MIMO communications significantly improve the achieved weighted sum rate. 
\end{abstract}
\section{Introduction}
Recently, \starriss have been proposed as a promising solution to support the high capacity of the coming \sixg communication systems. By employing a large number of low-cost STAR elements, the wireless signal impinging on STAR-RISs can be split into transmitted and reflected sides, thus realizing a full-space \sre~\cite{9424177}.\\
\indent In STAR-RIS assisted communications, a large STAR-RIS comprising of massive elements has to be employed for facilitating the considerable passive beamforming gain~\cite{bjornson2020power}. Moreover, STAR-RISs are extremely beneficially to be deployed for assisting millimeter-wave/Terahertz (mmWave/THz) communications via creating \los links to overcome the blockage issue~\cite{8901159}. However, the large aperture of STAR-RISs and the potentially high operating frequencies inevitably cause the spherical propagation to be dominated. As a result, the \emm wave propagation in the vicinity of STAR-RISs has to be accurately modeled with \emph{spherical waves} instead of the simple \emph{parallel waves}. The near-field wireless channels have following promising characteristics.
\begin{itemize}
\item \textbf{Near-field Beamfocusing:} Owing to the different EM wave propagation model, the near-field channel contains both angle and distance information of users~\cite{9536436}. In the near field of the STAR-RIS, the signals transmitted/reflected by the STAR-RIS can be concentrate on a specific location. This enables the STAR-RIS to carry out precise interference management among multiple users.
\item \textbf{High-rank \los MIMO Channels:} The STAR-RIS can create \los channels for blocked users. In contrast, thanks to the spherical propagation in the near-field of the STAR-RIS, the near-field \los channel is rank-sufficient~\cite{miller2019waves}. Compared with the low-rank far-field LoS channel, the rank-sufficient near-field LoS channel can support more information streams for the multi-antenna users.
\end{itemize}

Despite the above attractive benefits, to the best of the authors' knowledge, the STAR-RIS aided near-field communications and the corresponding joint beamforming design have not been investigated, which motivates this work. The main contributions of this work are summarized as follows:
\begin{itemize}
\item We solve the formulated non-convex problem by the developed block coordinate descent (BCD)-based algorithm, where the active beamforming matrices at the BS and the TRCs at the STAR-RIS are alternately optimized. Given fixed STAR TRCs, the optimal active beamforming matrices are obtained by solving a convex problem.
\item We develop two algorithms for optimizing the STAR-RIS TRCs under given fixed active beamforming matrices. First, we propose the penalty-based iterative (PEN) algorithm to optimize TRCs by invoking the successive convex approximation (SCA) technique. Then, for STAR-RIS with massive elements, we develop the low-complexity element-wise iterative (ELE) algorithm to optimize TRCs by invoking the bisection method. 
\end{itemize}
\section{System Model and Problem Formulation}
\subsection{System Model}
\begin{figure}[t!]
    \begin{center}
        \includegraphics[width=3in]{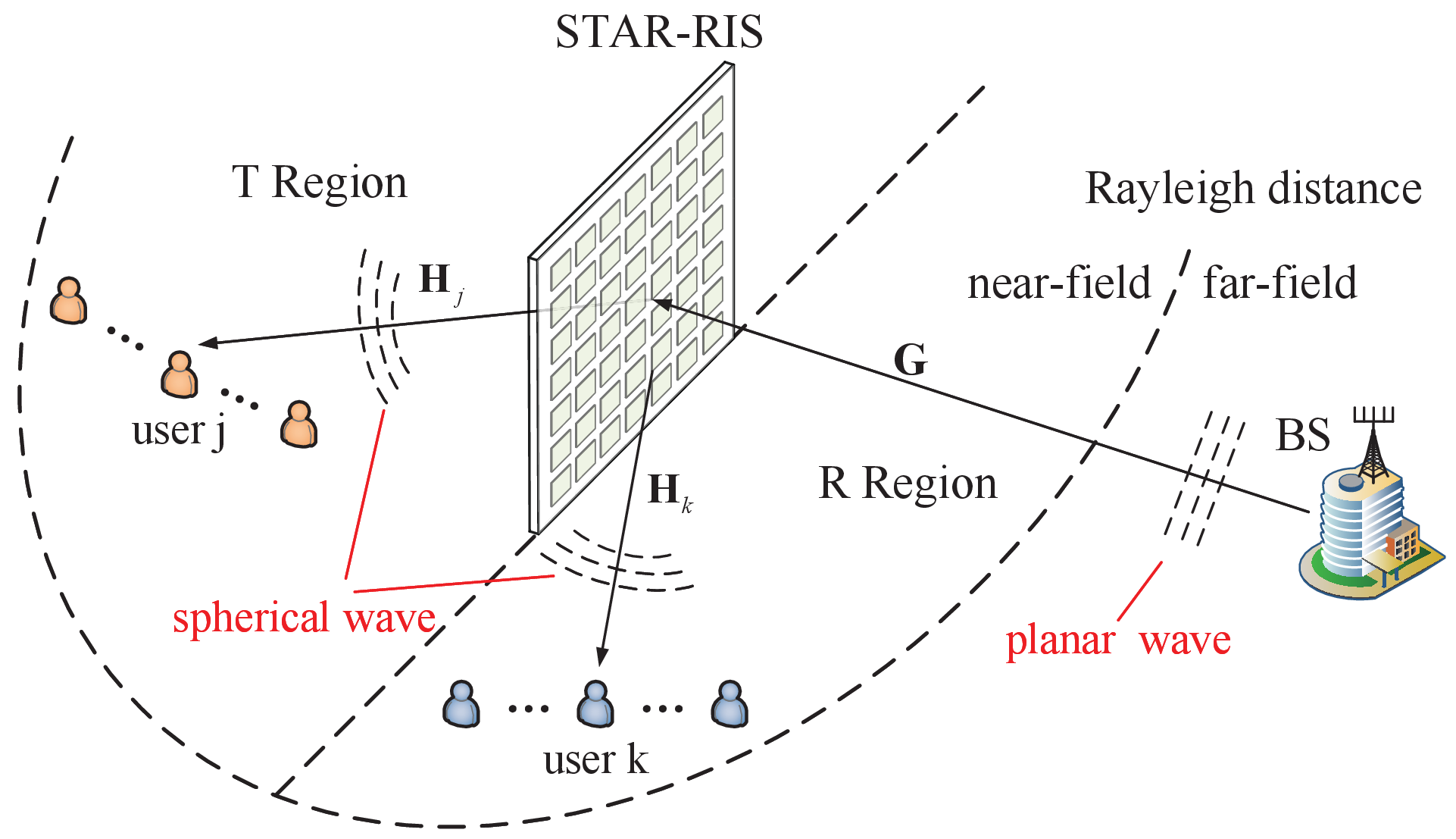}\vspace{-0.1cm}
        \caption{The Proposed System Model.}\vspace{-0.8cm}
        \label{system_model}
    \end{center}
\end{figure}
We consider a STAR-RIS aided downlink MIMO system, where an $M_b$-antenna BS serves multiple $M$-antenna users, whose indices are collected in $\mathcal{K}\buildrel \Delta \over = \left\{1,2, \cdots, K\right\}$. As depicted in Fig.~\ref{system_model}, the BS-user links are blocked by obstacles, which is relevant practical for systems operating under mmWave/THz bands. A STAR-RIS works under the energy splitting (ES) mode is deployed on the user side to create \los transmission and reflection links for the blocked users. It is assumed that the uniform planar array (UPA)-type STAR-RIS contains $N=N_y \times N_z$ elements, and the BS and users are equipped with uniform linear arrays (ULAs). The users locate in the transmission side and reflection side of the STAR-RIS are referred to as T users and R users, respectively. The indices of T users and R users are collected in subset $\mathcal{K}_t\buildrel \Delta \over = \left\{1,2, \cdots, K_0\right\}$ and subset $\mathcal{K}_r\buildrel \Delta \over = \mathcal{K}\setminus\mathcal{K}_t$. Let $\mathbf{\Phi}_{i}=\diag\{\sqrt{\rho_{1}^i}e^{j\theta_{1}^i}, \sqrt{\rho_{2}^i}e^{j\theta_{2}^i}, \cdots, \sqrt{\rho_{N}^i}e^{j\theta_{N}^i}\}, \forall i\in\{t,r\}$ denote the TRC matrix of the STAR-RIS, where $\rho_n^t, \rho_n^r\in[0,1]$ are the amplitude coefficients for transmission and reflection and $\theta_n^t, \theta_n^r\in[0,2\pi)$ are the corresponding phase shifts introduced by the $n$-th elements.
\subsection{Channel Model} 
As the STAR-RIS is deployed on the user side, the far-field channel model can be used for the BS-STAR-RIS link due to the relatively far distance, while the near-field channel model has to be used for the STAR-RIS-user link. Then, we introduce the far-field channel from the BS to the STAR-RIS and the near-field channel from the STAR-RIS to the user $k$, which are denoted by $\mathbf{G}\in\mathbb{C}^{N \times M_b}$ and $\mathbf{H}_k\in\mathbb{C}^{M \times N}$, respectively. 
\subsubsection{Far-field BS-STAR-RIS Channel}
For the far-field BS-STAR-RIS channel, we adopt the widely used geometric channel model with $L$ scatterers for the BS-STAR channel, which can be given by
\begin{equation}
\begin{aligned}\label{H}
\mathbf{G}=\sqrt{\frac{\beta M_b N}{L}} \sum\nolimits_{l=1}^{L} \mathbf{e}_\text{STAR}\left(\varphi_{l}, \vartheta_{l}\right)\mathbf{e}^{\mathrm{H}}_\text{BS}\left(\gamma_{l}\right) ,
\end{aligned}
\end{equation}
where $\gamma_{l}$ represents angle of departure (AoD) associated with the BS. $\varphi_{l}$ and $\vartheta_{l}$ represent azimuth angle of arrival (AoA) and elevation AoA associated with the STAR-RIS, respectively. $L$ is the number of dominant paths, $\beta$ is the path-loss coefficient. The array response vectors for the UPA-type STAR-RIS and the ULA at the BS are defined as in \cite{balanis2016antenna}.
\subsubsection{Near-field STAR-RIS-user Channels}
\begin{figure}[h]
\centering
\includegraphics[width=1.8in]{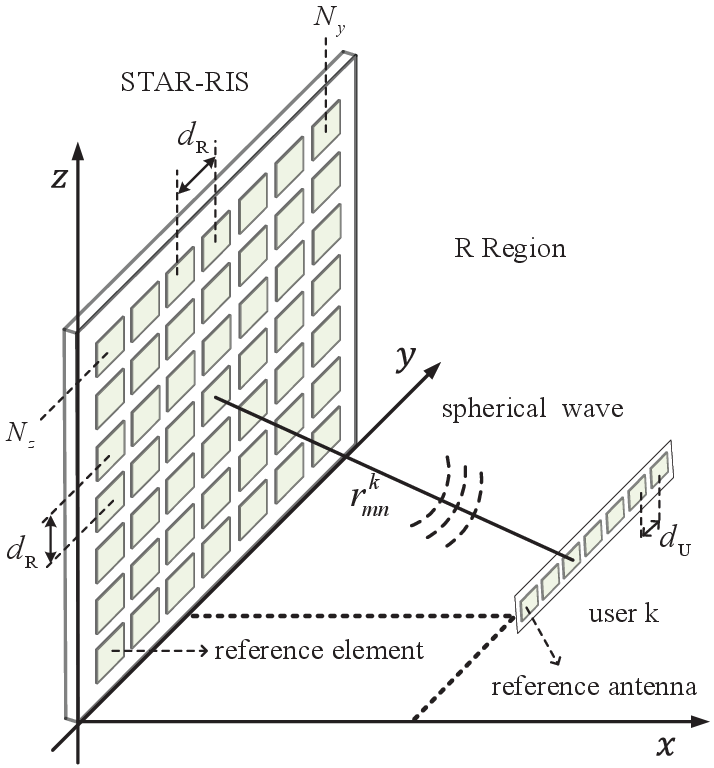}\vspace{-0.2cm}
\caption{The near-field STAR-RIS-user \los channel.}\vspace{-0.2cm}
\label{fig_2}
\end{figure}

Fig.~\ref{fig_2} shows the setup of the STAR-RIS and the user $k$, where the Cartesian coordinate of the reference element of the STAR-RIS and the reference antenna of the user $k$ can be denoted by $\left(0, y_f, z_f\right)$ and $\left(x_k, y_k, 0\right)$, respectively. The STAR-RIS is in the $YZ$-plane and all the users are in the $XY$-plane. If the STAR elements are indexed row by row from the bottom to the top, the Cartesian coordinate of the $n$-th STAR element is given by 
\begin{equation}
\mathbf{p}_{n}=\left[0, i_{\mathrm{y}}\left(n\right)d_R+y_f, i_{\mathrm{z}}\left(n\right)d_R+z_f\right]^{\mathrm{T}}, n\in\mathcal{N},
\end{equation}
where $\mathcal{N}\!=\!\left\{1,2, \cdots, N\right\}$. $d_R\!=\!\lambda_c$ is the STAR element spacing. $i_{\mathrm{y}}\left(n\right)\!=\!\mathrm{mod}\left(n-1,N_{\mathrm{y}}\right)$ and $i_{\mathrm{z}}\left(n\right)\!=\!\lfloor \left(n-1\right)/N_{\mathrm{y}}\rfloor$ are the column and row index of the $n$-th element, respectively.\\
\indent Furthermore, we assume that the ULAs of all users are parallel to the $y$-axis. Thus, the Cartesian coordinate of the $m$-th antenna of user $k$, $m\in\mathcal{M}=\{1,2,\cdots,M\}$, is given by
\begin{equation}
\mathbf{u}_{m}^k=\left[x_k, \left(m-1\right)d_U+y_z, 0\right]^{\mathrm{T}},
\end{equation}
where $d_U=\lambda_c/2$ is the user antenna spacing.

Then, the \los channel between the $n$-th STAR element and the $m$-th antenna of the user $k$ can be represented as
\begin{equation}
\begin{aligned}
\left[\mathbf{H}_k\right]_{mn}=\frac{\lambda_c}{4\pi \left\|\mathbf{u}_{m}^k-\mathbf{p}_{n}\right\|_2}\exp\left(-j2\pi \left\|\mathbf{u}_{m}^k-\mathbf{p}_{n}\right\|_2/\lambda_c\right).
\end{aligned}
\end{equation}
\subsection{Communication Model}
The signal sent by the BS can be expressed as
\begin{equation}
\begin{aligned}
\mathbf{x}=\sum\nolimits_{k\in\mathcal{K}}\mathbf{W}_k\mathbf{s}_k,
\end{aligned}
\end{equation}
where $\mathbf{s}_k\in\mathbb{C}^{M \times 1}$ and $\mathbf{W}_k\in\mathbb{C}^{M_b \times M}$ represent the symbol vector and the beamformer for user $k$, respectively.\\
\indent Then, the received signal vector at the user $k$, $\forall k \in \mathcal{K}_i$, $\forall i \in \{t,r\}$, is given by
\begin{equation}\label{y}
\begin{aligned}
\mathbf{y}_k=\mathbf{H}_k\mathbf{\Phi}_i\mathbf{G}\mathbf{W}_k\mathbf{s}_k+\sum\nolimits_{l\in\mathcal{K}\setminus k}\mathbf{H}_k\mathbf{\Phi}_i\mathbf{G}\mathbf{W}_{l}\mathbf{s}_{l}+\mathbf{n}_k,
\end{aligned}
\end{equation}
where $\mathbf{n}_k\in\mathbb{C}^{M \times 1}$ is the noise vector that satisfies $\mathcal{CN}\left(\mathbf{0}, \sigma^2\mathbf{I}\right)$. $\sigma^2$ is the noise power.\\
\indent The achievable data rate of the user $k$, $\forall k \in \mathcal{K}_i$, $\forall i \in \{t,r\}$, is given by
\begin{equation}\label{R}
\begin{aligned}
R_k\left(\mathbf{W},\mathbf{\Phi}\right)\!=\!\log\left|\mathbf{I}+\mathbf{H}_k\mathbf{\Phi}_i\mathbf{G}\mathbf{W}_k\mathbf{W}_k^\mathrm{H}\mathbf{G}^\mathrm{H}\mathbf{\Phi}_i^\mathrm{H}\mathbf{H}_k^\mathrm{H}\mathbf{J}_k^{-1}\right|,
\end{aligned}
\end{equation}
where we have $\mathbf{W}= \{\mathbf{W}_k, \forall k\in\mathcal{K}\}$, $\mathbf{\Phi} = \{\mathbf{\Phi}_t,\mathbf{\Phi}_r\}$, and $\mathbf{J}_k$ is the interference-plus-noise covariance matrix
\begin{equation}\label{J}
\begin{aligned}
\mathbf{J}_k=\sum\nolimits_{l\in\mathcal{K}\setminus k}\mathbf{H}_k\mathbf{\Phi}_i\mathbf{G}\mathbf{W}_{l}\mathbf{W}_{l}^\mathrm{H}\mathbf{G}^\mathrm{H}\mathbf{\Phi}_i^\mathrm{H}\mathbf{H}_k^\mathrm{H}+\sigma^2\mathbf{I}.
\end{aligned}
\end{equation}
\subsection{Optimization Problem Formulation}
The weighted sum rate maximization problem can be formulated as follows
\begin{subequations}\label{P1}
\begin{align}
\label{OBP1}\mathop {\min }\limits_{ {{\mathbf{W}},{\mathbf{\Phi}}} } \;\;&\sum\nolimits_{k\in\mathcal{K}}\eta_k R_k\left(\mathbf{W},\mathbf{\Phi}\right)\\
\label{Phi}{\rm{s.t.}}\;\;& \rho_n^t,\rho_n^r\ge 0, \rho_n^t+\rho_n^r=1, \forall n \in \mathcal{N},\\
\label{power}&\sum\nolimits_{k\in\mathcal{K}}\|\mathbf{W}_k\|_{\text{F}}^{2}\le P,
\end{align}
\end{subequations}
where the constraint \eqref{Phi} is the TRC constraints for the STAR elements. The constraint \eqref{power} is the power constraint of the BS, with $P$ denoting the maximum transmit power of the BS. $\eta_k$ is the factor denotes the access priority for user $k$. Then, we solve problem \eqref{P1} using the block coordinate descent (BCD) method.
\section{Proposed BCD Algorithm}
\subsection{Problem Reformulation}
First, we use the weighted minimum mean square error (WMMSE) method to convert \eqref{OBP1} into a more tractable form. Assume that the receivers adopt linear combining matrices, then the combined signal vector of the user $k\in\mathcal{K}$ is given by
\begin{equation}
\begin{aligned}
\tilde{\mathbf{s}}_k=\mathbf{U}_k^\mathrm{H}\mathbf{y}_k,
\end{aligned}
\end{equation}
where ${\mathbf{U}}_k\in\mathbb{C}^{M \times M}$ is the combining matrix for the user $k$.\\
\indent Then, the MSE matrix of the user $k\in\mathcal{K}$ is given by 
\begin{equation}
\begin{aligned}
{\mathbf{E}}_k=&\left(\mathbf{U}_k^\mathrm{H}\bar{\mathbf{H}}_k\mathbf{W}_k-\mathbf{I}\right)\left(\mathbf{U}_k^\mathrm{H}\bar{\mathbf{H}}_k\mathbf{W}_k-\mathbf{I}\right)^\mathrm{H}\\
&+\mathbf{U}_k^\mathrm{H}\bar{\mathbf{H}}_k\mathbf{W}_{\bar{k}}\mathbf{W}_{\bar{k}}^\mathrm{H}\bar{\mathbf{H}}_k^\mathrm{H}\mathbf{U}_k+\sigma^2\mathbf{U}_k^\mathrm{H}\mathbf{U}_k,
\end{aligned}
\end{equation} 
where $\bar{\mathbf{H}}_k\in\mathbb{C}^{M \times M_b}={\mathbf{H}}_k{\mathbf{\Phi}}_i{\mathbf{G}}$ is the aggregation channel between the BS and the user $k, \forall k \in \mathcal{K}_i$, $\forall i \in \{t,r\}$.\\
\indent Then, problem \eqref{P1} can be reformulated as follows:
\begin{subequations}\label{P2}
\begin{align}
\label{OBP2}\mathop {\max }\limits_{ {{\mathbf{Z}},{\mathbf{U}},{\mathbf{W}},{\mathbf{\Phi}}} } \;\;&\sum\nolimits_{k\in\mathcal{K}}\eta_k\left(\log\left|\mathbf{Z}_k\right|-\tr\left(\mathbf{Z}_k\mathbf{E}_k\right)+M\right)\\[-0.1cm]
{\rm{s.t.}}\;\;& \eqref{Phi},\eqref{power},
\end{align}
\end{subequations}
where $\mathbf{Z}=\{\mathbf{Z}_k\succeq \mathbf{0}, \forall k \in \mathcal{K}\}$ and $\mathbf{U}=\{\mathbf{U}_k, \forall k \in \mathcal{K}\}$ denote the set of auxiliary matrices and the set of combining matrices, respectively.
\subsection{Proposed BCD algorithm for joint beamforming }
In the following, we solve the problem \eqref{P2} iteratively using the BCD method.
\subsubsection{Subproblem with respect to $\mathbf{U}$}
Given $\mathbf{Z}$, $\mathbf{W}$ and $\mathbf{\Phi}$, the optimal $\mathbf{U}$ in \eqref{P2} is the solution of ${{\partial {f_{k}}} \mathord{\left/
{\vphantom {{\partial {f_{k}}} {\partial {\mathbf{U}_k}}}} \right.\kern-\nulldelimiterspace} {\partial {\mathbf{U}_k}}}=\mathbf{0}$, for $\forall k \in \mathcal{K}$. 
\subsubsection{Subproblem with respect to $\mathbf{Z}$}
Given $\mathbf{U}$, $\mathbf{W}$ and $\mathbf{\Phi}$, the optimal $\mathbf{Z}$ in \eqref{P2} can be obtained by solving ${{\partial {f_{k}}} \mathord{\left/
{\vphantom {{\partial {f_{k}}} {\partial {\mathbf{Z}_k}}}} \right.\kern-\nulldelimiterspace} {\partial {\mathbf{Z}_k}}}=\mathbf{0}$, for $ \forall k \in \mathcal{K}$. The solution can be given by $\mathbf{Z}_k={\mathbf{E}}_k^{-1}$.
\subsubsection{Subproblem with respect to $\mathbf{W}$}
Furthermore, given $\mathbf{Z}$, $\mathbf{U}$ and $\mathbf{\Phi}$, the weighted MSE  minimization problem \eqref{P2} can be transformed into the active beamforming problem as
\begin{subequations}\label{P3}
\begin{align}
\label{OBP3}\mathop {\min }\limits_{ {{\mathbf{W}}} } \;\;&\sum\nolimits_{k\in\mathcal{K}}\tr\left(\mathbf{W}_k^\mathrm{H}\mathbf{A}\mathbf{W}_k-2\tr\left(\Re\left(\mathbf{B}_k\mathbf{W}_k\right)\right)\right)\\[-0.1cm]
{\rm{s.t.}}\;\;& \eqref{power},
\end{align}
\end{subequations}
where $\mathbf{A}$ and $\mathbf{B}_k$ $\forall k \in \mathcal{K}$ are denoted by
\begin{equation}
\begin{aligned}
\mathbf{A}=\sum\nolimits_{l\in\mathcal{K}}\eta_l\bar{\mathbf{H}}_l^\mathrm{H}\mathbf{U}_l\mathbf{Z}_l\mathbf{U}_l^\mathrm{H}\bar{\mathbf{H}}_l, \quad \mathbf{B}_k=\eta_k\mathbf{Z}_k\mathbf{U}_k^\mathrm{H}\bar{\mathbf{H}}_k.
\end{aligned}
\end{equation}

Problem~\eqref{P3} is a quadratically constrained quadratic program problem and can be solved by CVX~\cite{grant2014cvx}. To reduce the complexity, the closed form solution to the transmit beamformers is given in~\cite{pan2020multicell} and its computational complexity is given by $\mathcal{O}\left(K M_b^3\right)$.
\subsubsection{Subproblem with respect to $\mathbf{\Phi}$}
Finally, for given $\mathbf{Z}$, $\mathbf{U}$ and $\mathbf{W}$, substitute $\mathbf{E}_k$ into \eqref{OBP2} and ignoring the terms that are not related to $\mathbf{\Phi}$. Then the TRC matrices optimization problem is formulated as
\begin{subequations}\label{P4}
\begin{align}
\label{OBP4}\mathop {\min }\limits_{ {{\mathbf{\Phi}}} } \;\;&\sum\nolimits_{k\in\mathcal{K}}\tr\left(\mathbf{\Phi}_i^\mathrm{H}\mathbf{C}_k\mathbf{\Phi}_i\mathbf{D}-2\Re\left(\mathbf{E}_k\mathbf{\Phi}_i\right)\right)\\[-0.1cm]
{\rm{s.t.}}\;\;& \eqref{Phi},
\end{align}
\end{subequations}
wherein $\mathbf{C}_k$, $\mathbf{D}$, and $\mathbf{E}_k$ are denoted by $\mathbf{C}_k=\eta_k\mathbf{H}_k^\mathrm{H}\mathbf{U}_k\mathbf{Z}_k\mathbf{U}_k^\mathrm{H}\mathbf{H}_k$, $\mathbf{D}=\mathbf{G}\left(\sum_{l\in\mathcal{K}}\mathbf{W}_l\mathbf{W}_l^\mathrm{H}\right)\mathbf{G}^\mathrm{H}$, and $\mathbf{E}_k=\eta_k\mathbf{G}\mathbf{W}_k\mathbf{Z}_k\mathbf{U}_k^\mathrm{H}{\mathbf{H}}_k, \forall k \in \mathcal{K}$, respectively.
\subsection{Proposed PEN Algorithm}
Using the matrix identities, $\forall k \in \mathcal{K}_i, \forall i \in \{t,r\}$,
\begin{equation}
\begin{aligned}
\tr\left(\mathbf{\Phi}_i^\mathrm{H}\mathbf{C}_k\mathbf{\Phi}_i\mathbf{D}\right)=\mathbf{v}_i^\mathrm{H}\mathbf{F}_k\mathbf{v}_i, \quad \tr\left(\mathbf{\Phi}_i\mathbf{E}_k\right)=\mathbf{e}_k^\mathrm{T}\mathbf{v}_i,
\end{aligned}
\end{equation}
where $\mathbf{v}_{i}\!=\!\Diag\left(\mathbf{\Phi}_{i}\right)\!=\!\left[v^i_1, v^i_2, \cdots, v^i_N\right]^{\mathrm{T}}$, $\mathbf{F}_k=\left(\mathbf{C}_k \odot \mathbf{D}^\mathrm{T}\right)$ and $\mathbf{e}_{k}=\Diag\left(\mathbf{E}_{k}\right)$, \eqref{OBP4} can be reformulated as
\begin{equation}\label{Gk}
\begin{aligned}
\sum\nolimits_{i\in \{t,r\}}\left(\mathbf{v}_i^\mathrm{H}\mathbf{F}_i\mathbf{v}_i-2\Re\left(\mathbf{e}_i^\mathrm{T}\mathbf{v}_i\right)\right),
\end{aligned}
\end{equation}
where $\mathbf{v} = \left\{\mathbf{v}_t, \mathbf{v}_r\right\}$, $\mathbf{F}_i=\sum_{k\in\mathcal{K}_i}\mathbf{F}_k$ and $\mathbf{e}_i=\sum_{k\in\mathcal{K}_i}\mathbf{e}_k$.\\
\indent To facilitate the design, we define the augmented TRC vector $\bar{\mathbf{v}}_i=\left[\mathbf{v}_i^\mathrm{T} \ \sqrt{t_i}\right]^\mathrm{T}, \forall i\in\{t,r\}$, where $\{t_t, t_r\}$ $\in\mathbb{R}$ are auxiliary variables. Moreover, we define ${\mathbf{V}}_i=\bar{\mathbf{v}}_i\bar{\mathbf{v}}_i^\mathrm{H}$, which satisfies ${\mathbf{V}}_i\succeq 0$, $\rank\{{\mathbf{V}}\}_i=1$, and $\Diag\{{\mathbf{V}}_i\}={\bm{\rho}}_i$, where ${\bm{\rho}}_i=\left[\rho_{1}^i,\rho_{2}^i,\cdots,\rho_{N}^i,t^i\right]^\mathrm{T}$. We rewrite the quadratic term $\mathbf{v}_i^{\mathrm{H}}\mathbf{F}_i\mathbf{v}_i-2\Re\{\mathbf{e}_i^{\mathrm{T}}\mathbf{v}_i\}$ as $\tr\left({\mathbf{V}}_i\bar{\mathbf{F}}_i\right)$ with $\bar{\mathbf{F}}_i$ defined as\vspace{-0.1cm}
\begin{equation}
\begin{aligned}
\bar{\mathbf{F}}_i=\begin{bmatrix}
\mathbf{F}_i & -\mathbf{e}_i^{*} \\
-\mathbf{e}_i^\mathrm{T} & 0 
\end{bmatrix}, \forall i \in \{t,r\}.
\end{aligned}\vspace{-0.1cm}
\end{equation}

Then, the problem \eqref{P4} can be equivalently written as \vspace{-0.1cm}
\begin{subequations}\label{P5}
\begin{align}
&\mathop {\min }\limits_{ {{\mathbf{V}},{\bm{\rho}}} } \;\;\sum\nolimits_{\forall i \in \{t,r\}}\tr\left({\mathbf{V}}_i\bar{\mathbf{F}}_i\right)  \\[-0.1cm]
\label{rank V}{\rm{s.t.}}\;\;&\rank\{{\mathbf{V}}_i\}=1, \forall i \in\{t,r\},\\[-0.1cm]
\label{amplitude1}&\Diag\{{\mathbf{V}}_i\}={\bm{\rho}}_i, \forall i \in\{t,r\},\\[-0.1cm]
\label{semi V}&{\mathbf{V}}_i\succeq 0, \forall i \in\{t,r\},\\[-0.1cm]
\label{amplitude2}&\rho_n^t, \rho_n^r \in [0,1], \rho_n^t+\rho_n^r=1, \forall n\in \mathcal{N},\\[-0.1cm]
\label{auxiliary}&t^i=1, \forall i \in\{t,r\},
\end{align}
\vspace{-0.4cm}
\end{subequations} 

\noindent where $\mathbf{V}=\{{\mathbf{V}}_t,{\mathbf{V}}_r\}$ and $\bm{\rho} =\{{\bm{\rho}}_t,{\bm{\rho}}_r\}$.

To tackle the non-convex rank-one constraint~\eqref{rank V}, we rewrite it as the following equality constraint
\begin{equation}\label{nuclear}
\begin{aligned}
\left\|{\mathbf{V}}_i\right\|_*-\left\|{\mathbf{V}}_i\right\|_2=0, \forall i\in\{t,r\},
\end{aligned}
\end{equation}
where $\left\|{\mathbf{V}}_i\right\|_*$ and $\left\|{\mathbf{V}}_i\right\|_2$ denote the nuclear norm and the spectral norm of ${\mathbf{V}}_i$, respectively. Note that for any ${\mathbf{V}}_i\in\mathbb{H}^{N+1}$ and ${\mathbf{V}}_i \succeq 0$, the equality constraint in~\eqref{nuclear} is met if and only if the constraint~\eqref{rank V} is met.\\
\indent Next, we obtain the following optimization problem by exploiting the penalty method and incorporating the reformulated rank-one constraint~\eqref{nuclear} into the objective function
\begin{subequations}\label{P6}
\begin{align}
\label{P6O} \mathop {\min }\limits_{ {{\mathbf{V}},{\bm{\rho}}} } \;\;&\sum\nolimits_{ \forall i \in \{t,r\}}\left(\tr\left({\mathbf{V}}_i\bar{\mathbf{F}}_i\right)+\mu\left(\left\|{\mathbf{V}}_i\right\|_*-\left\|{\mathbf{V}}_i\right\|_2\right)\right)\\
{\rm{s.t.}}\;\;& \eqref{amplitude1}-\eqref{auxiliary},
\end{align}
\end{subequations}
where $\mu>0$ denotes the penalty factor. The problem~\eqref{P6} is non-convex since its objective function is non-convex.\\
\indent Then, we adopt the SCA technique to solve problem~\eqref{P6} iteratively with the first-order Taylor expansion. For any feasible point $\mathbf{V}^{\left(n\right)}\buildrel \Delta \over=\{\mathbf{V}^{\left(n\right)}_t,\mathbf{V}^{\left(n\right)}_r\}$ in the $n$-th SCA iteration, a convex upper bound of the term $\left\|{\mathbf{V}}_i\right\|_*-\left\|{\mathbf{V}}_i\right\|_2$ can be given by 
\begin{equation}\label{SCA}
\begin{aligned}
f_{\text{SCA}}&\left({\mathbf{V}}_i,\mathbf{V}^{\left(n\right)}_i\right)=\left\|{\mathbf{V}}_i\right\|_*-\|\mathbf{V}^{\left(n\right)}_i\|_2-\\
&{\mathbf{d}_{\text{max}}^{(n)}}^\mathrm{H}\left(\mathbf{V}_i-\mathbf{V}^{\left(n\right)}_i\right)\mathbf{d}_{\text{max}}^{(n)}\ge\left\|{\mathbf{V}}_i\right\|_*-\left\|{\mathbf{V}}_i\right\|_2
\end{aligned}
\end{equation}
where $\mathbf{d}_{\text{max}}^{(n)}$ denotes the eigenvector associated with the largest eigenvalue of $\mathbf{V}^{\left(n\right)}_k$.\\
\indent For the $n$-th SCA iteration, a convex semidefinite program (SDP) problem can be obtained by substituting~\eqref{SCA} into problem~\eqref{P6}
\begin{subequations}\label{P7}
\begin{align}
\label{OBP7}\mathop {\min }\limits_{ {{\mathbf{V}},{\bm{\rho}}} } \;\;&\sum\nolimits_{ \forall i \in \{t,r\}}\tr\left({\mathbf{V}}_i\bar{\mathbf{F}}_i\right)+\mu f_{\text{SCA}}\left({\mathbf{V}}_i,\mathbf{V}^{\left(n\right)}_i\right)\\
{\rm{s.t.}}\;\;& \eqref{amplitude1}-\eqref{auxiliary}.
\end{align}
\end{subequations}

The propose PEN algorithm for ES is summarized in \textbf{Algorithm 1}. The main complexity of Algorithm~\ref{algorithm1} is caused by solving the SDP problem~\eqref{P7} and can be approximated as $\mathcal{O}\left(N^3\right)$. So the computational complexity of Algorithm~\ref{algorithm1} is given by $\mathcal{O}\left(\emph{I}_\text{out} \emph{I}_\text{in} N^3\right)$, where $\emph{I}_\text{out}$ and $\emph{I}_\text{in}$ are the numbers of iterations for the outer loop and the inner loop, respectively.

\begin{algorithm}[htbp]
\caption{Proposed PEN Algorithm for Solving Problem~\eqref{P5}}\label{algorithm1}
\begin{algorithmic}[1]
\STATE {Initialize feasible points $\mathbf{V}^{(0)}$, the penalty factor $\mu$.}
\STATE {\bf repeat: outer loop}
\STATE \quad Set inner loop index $n = 0$.
\STATE \quad {\bf repeat: inner loop}
\STATE \quad\quad Get the intermediate solution by solve the problem~\eqref{P7} for given $\mathbf{V}^{(n)}$.
\STATE \quad\quad Update $\mathbf{V}^{(n+1)}$ with the intermediate solution and $n=n+1$.
\STATE \quad {\bf until} the fractional decrease of~\eqref{OBP7} is less than the threshold $\epsilon_{\text{SCA}}$.
\STATE \quad Get the solution of problem~\eqref{P6} with penalty factor $\mu$.
\STATE \quad Update $\mathbf{V}^{(0)}$ with the current solution $\mathbf{V}^{(n)}$.
\STATE \quad Update the penalty factor as $\mu=\omega\mu$.
\STATE {\bf until} the constraint violation is less than the threshold $\epsilon_{\text{p}} >0$.
\end{algorithmic}
\end{algorithm}
\subsection{Proposed ELE Algorithm}
The STAR-RIS requires massive elements for facilitating the considerable passive beamforming gain. When the number of STAR elements is large, \eg, several hundreds of or even thousands of elements, the computational complexity involved in optimizing TRCs using the PEN algorithm becomes impractical to implement. To address this problem, the low-complexity ELE algorithm is proposed to solve problem \eqref{P4} in the following. Specifically, the TRCs of each STAR element are optimized in turn when other $N-1$ elements fixed. 

Define $f^i_{qj}$ as the element in the $q$-th row and the $j$-th column of matrix $\mathbf{F}_i\in\mathbb{C}^{N\times N}$ and let $e_n^i$ and $v_n^i$ denote the $n$-th element of the vector $\mathbf{e}_i\in\mathbb{C}^{N\times 1}$ and $\mathbf{v}_i\in\mathbb{C}^{N\times 1}$, respectively. Recall that $\mathbf{v}_{i}=\diag\left(\mathbf{\Phi}_{i}\right)$, the term $\mathbf{e}_i^\mathrm{T}\mathbf{v}_i, \forall i\in\{t,r\}$ in \eqref{Gk} can be written as 
\begin{equation}\label{e_expand}
\begin{aligned}
\mathbf{e}_i^\mathrm{T}\mathbf{v}_i= e_n^i v_n^{i} + \sum\nolimits_{j\neq n}^N e_{j}^i v_j^{i}.
\end{aligned}
\end{equation}

And the term $\mathbf{v}_i^\mathrm{H}\mathbf{F}_i\mathbf{v}_i, \forall i\in\{t,r\}$ in \eqref{Gk} can be written as
\begin{equation}\label{F_expand}
\begin{aligned}
\left|v_n^{i}\right|^2 f_{n,n}^i + 2\Re\left(\sum\nolimits_{q\neq n}^N (v_q^{i})^* f_{q,n}^i v_n^{i}\right) + \sum\nolimits_{q,j\neq n}^N (v_q^{i})^* f_{q,j}^iv_j^{i}.
\end{aligned}
\end{equation}

Substituting \eqref{e_expand} and \eqref{F_expand} into \eqref{Gk}, $\sum_{k\in\mathcal{K}_i}g_k\left(\mathbf{v}\right)$ can be rewritten as
\begin{equation}\label{Gk_expand}
\begin{aligned}
\sum\nolimits_{k\in\mathcal{K}_i}g_k\left(\mathbf{v}\right)=\left|v_n^{i}\right|^2A_n^i + 2\Re\left(B_n^iv_n^{i}\right) + C_n^i.
\end{aligned}
\end{equation}
where
\begin{equation}
\begin{aligned}
&A_n^{i}=f_{n,n}^i, \quad B_n^i=\sum\nolimits_{q\neq n}^N (v_q^{i})^* f_{q,n}^i-e_n^i,\\[-0.1cm]
&C_n^{i}=\sum\nolimits_{q,j\neq n}^N (v_q^{i})^* f_{q,j}^i v_j^{i}-2\Re\left(\sum\nolimits_{j\neq n}^N e_{j}^iv_j^{i}\right).
\end{aligned}
\end{equation}

For given TRCs for $N-1$ STAR elements, the design of the $n$-th transmission and reflection coefficients can be formulated as the following problem 
\begin{subequations}\label{P10}
\begin{align}
\mathop {\min }\limits_{ {{v_n^t},{v_n^r}} } \;\;&\sum\nolimits_{ \forall i \in \{t,r\}}\left|v_n^{i}\right|^2A_n^i + 2\Re\left(B_n^iv_n^{i}\right)\\
{\rm{s.t.}}\;\;& \left|v_n^t\right|+\left|v_n^r\right|=1.
\end{align}
\end{subequations}

The optimal phases of the $n$-th STAR element, \ie, $\theta_n^i, \forall i \in \{t,r\}$ can be given by 
\begin{equation}\label{phaseopt}
\begin{aligned}
\theta_n^i = \left\{ {\begin{array}{*{20}{l}}
{\pi  - \angle B_n^i, B_n^i \in[0,\pi)}\\[-0.1cm]
{3\pi  - \angle B_n^i, B_n^i \in[\pi,2\pi)}
\end{array}} \right. 
\end{aligned}
\end{equation}

Subtitling \eqref{phaseopt} into problem \eqref{P10}, the optimization problem for the amplitudes of the $n$-th STAR element can be given by
\begin{subequations}\label{P12}
\begin{align}
\mathop {\min }\limits_{ {{\rho_n^t},{\rho_n^r}} } \;\;&\sum\nolimits_{ \forall i \in \{t,r\}}\rho_n^iA_n^i - 2\sqrt{\rho_n^i}\left|B_n^i\right|\\[-0.1cm]
{\rm{s.t.}}\;\;& \rho_n^t+\rho_n^r=1.
\end{align}
\end{subequations}

Using $\rho_n^r=1-\rho_n^t$, problem \eqref{P12} can be reformulated as an unconstrained optimization problem 
\begin{equation}\label{P13}
\begin{aligned}
\mathop {\min }\nolimits_{ {\rho_n^t} } \left(A_n^t-A_n^r\right)\rho_n^t-2\sqrt{\rho_n^t}\left|B_n^t\right|-2\sqrt{1-\rho_n^t}\left|B_n^r\right|.
\end{aligned}
\end{equation}

It can be easily verified that problem \eqref{P13} is convex with respect to $\rho_n^t$. Then the solution to problem \eqref{P13} (as well as the solution to problem problem \eqref{P12}) can be obtained by solving ${{d {f_{n}}} \mathord{\left/
{\vphantom {{d {f_{n}}} {d {\rho_n^t}}}} \right.\kern-\nulldelimiterspace} {d {\rho_n^t}}}=0$, where $f_{n}\left(\rho_n^t\right)$ denotes the objective function of problem \eqref{P13}. It is plausible that $f_{n}'\left(\rho_n^t\right)$ is a monotonically increasing function for $\rho_n^t$ and satisfies
\begin{equation}
\begin{aligned}
\mathop {\lim }\limits_{\rho_n^t \to 0^+}f_{n}'\left(\rho_n^t\right) \to -\infty, \mathop {\lim }\limits_{\rho_n^t \to 1^-}f_{n}'\left(\rho_n^t\right) \to \infty. 
\end{aligned}
\end{equation}

Then the solution of ${{d {f_{n}}} \mathord{\left/
{\vphantom {{d {f_{n}}} {d {\rho_n^t}}}} \right.\kern-\nulldelimiterspace} {d {\rho_n^t}}}=0$, \ie, $\left(\rho_n^t\right)_{\text{opt}}$, can be efficiently obtained by using the bisection search algorithm. The number of iterations for the bisection search algorithm to converge is given by $\log_2\left(\frac{1}{\epsilon}\right)$, where $\epsilon$ is convergence threshold for bisection. From the analysis, it is found that the computational complexity for updating STAR-RIS TRCs linearly increases with the number of STAR elements, i.e., $N$.
\subsection{The Overall Algorithm to Solve Problem~\eqref{P2}}
The overall BCD algorithm for solving problem~\eqref{P2} is summarized in \textbf{Algorithm 2}. The main complexity for the proposed BCD-PEN algorithm comes from solving the problem~\eqref{P3} with Lagrangian multiplier method and solving the problem~\eqref{P5} with Algorithm~\ref{algorithm1}. The main complexity for the proposed low-complexity BCD-ELE algorithm comes from updating the transmit beamforming matrices using the Lagrangian multiplier method and updating the TRC matrices in a element-wise manner. So the aggregated computational complexity of the BCD-PEN algorithm and the BCD-ELE algorithm can be given by $\mathcal{O}\left({I}_\text{BCD}\left(K M_b^3 + {I}_\text{out} {I}_\text{in}  N^3\right)\right)$ and $\mathcal{O}\left(\emph{I}_\text{BCD}\left(K M_b^3 +  N\right)\right)$, respectively. ${I}_\text{BCD}$ is the number of iterations.
\begin{algorithm}[htbp]
\caption{Proposed BCD Algorithm for Solving Problem~\eqref{P2}}\label{algorithm2}
\begin{algorithmic}[1]
\STATE {Initialize feasible $\mathbf{\Phi}$ and $\mathbf{W}$ that satisfy~\eqref{Phi} and~\eqref{power}.}
\STATE {{\bf{repeat:}}}
\STATE {\quad Given $\mathbf{W}$ and $\mathbf{\Phi}$, update the combining matrices $\mathbf{U}$.}
\STATE {\quad Given $\mathbf{W}$, $\mathbf{\Phi}$ and $\mathbf{U}$, update the auxiliary matrices $\mathbf{Z}$.}
\STATE {\quad Given $\mathbf{\Phi}$, $\mathbf{U}$ and $\mathbf{Z}$, update the transmit beamforming matrices $\mathbf{W}$ by solving problem~\eqref{P3}.}
\STATE {{\bf{if}} the BCD-PEN algorithm is adopted {\bf{then}}}
\STATE {\quad Given $\mathbf{W}$, $\mathbf{U}$ and $\mathbf{Z}$, update the TRC matrices $\mathbf{\Phi}$ by solving problem~\eqref{P5} with Algorithm~\ref{algorithm1}.}
\STATE {{\bf{else if}} the BCD-ELE algorithm is adopted {\bf{then}}}
\STATE {\quad{\bf{for}} $n\in\mathcal{N}$\bf{:}}
\STATE {\quad\quad Given $\mathbf{W}$, $\mathbf{U}$, $\mathbf{Z}$ and TRCs for other $N-1$ elements, update $v_n^t$ and $v_n^r$ using~\eqref{phaseopt} and the bisection algorithm.}
\STATE {\quad{\bf{end for}}}
\STATE {{\bf{end if}}}
\STATE  {\bf until} the fractional increase of the objective function
value of problem~\eqref{P2} is below a predefined threshold $\epsilon_{\text{BCD}}$.
\end{algorithmic}
\end{algorithm} 
\section{Simulation Results}
\subsection{Simulation Setup}
As shown in Fig.~\ref{user_setup}, the BS and the STAR-RIS are deployed at $\left({0, 0, 0}\right)$ m and $\left({0, 50, 0}\right)$ m, respectively. All users lie on the circles surrounding the STAR-RIS with a radius of $d_t^1=d_r^1=2$~m or with a radius of $d_t^2=d_r^2=4$~m. The angles of the BS-STAR channel paths are randomly generated following the uniform distribution. Channel path-loss coefficient is modeled as $\beta=C_{0}\left(\frac{d}{D_{0}}\right)^{-\alpha}$, where $C_{0}=-30$ dB is the path loss at reference distance $D_{0}=1$ m. $d$ denotes the distances between the BS and the STAR-RIS. $\alpha=2.2$ denotes path loss exponent of the BS-STAR-RIS channel. Other system parameters are set as follows: $f_c=10$ GHz, $\lambda_c=0.03$ m, $L=16$, $\sigma^{2}=-110$ dBm, $M_b=16$, $M=4$, $K=4$ with $\mathcal{K}_t=\left\{1,2\right\}$ and $\mathcal{K}_r=\left\{3,4\right\}$, weighting factors $ \eta_k=1, \forall k$. The Rayleigh distance of a UPA-type STAR-RIS with $40=5\times 8$ elements operating at $10$ GHz is about $5$ m. This indicates that all users lie in the near-field of the STAR-RIS.
\subsection{Baseline Schemes and User Setups}
We compare the proposed STAR-RIS near-field beamforming algorithms with the following three baseline schemes.
\begin{enumerate}
\item \textbf{Conventional RIS:} This baseline scheme utilizes a conventional reflecting RIS and a transmitting RIS. Each conventional RIS is assumed to possess $N/2$ elements.
\item \textbf{Uniform Energy Splitting STAR-RIS:} All the elements of the ES STAR-RIS utilize equal amplitude coefficients for transmission and reflection, respectively.
\item \textbf{Far-field \los channel-based beamforming:} This baseline scheme adopts the far-field \los channel model to represent the STAR-user channels.\\\vspace{-0.3cm}
\end{enumerate}
\begin{figure}[!htbp]\vspace{-0.3cm}
\centering
\subfigure[Random User.]{\label{Random_user}
\includegraphics[width= 2.3in]{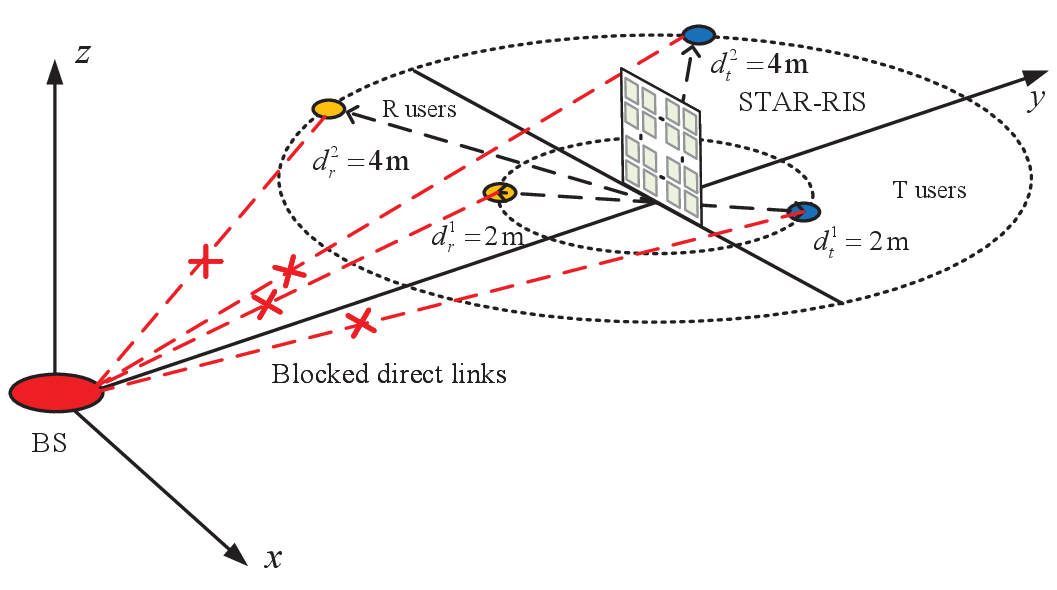}}
\subfigure[Inline User.]{\label{Inline_user}
\includegraphics[width= 2.3in]{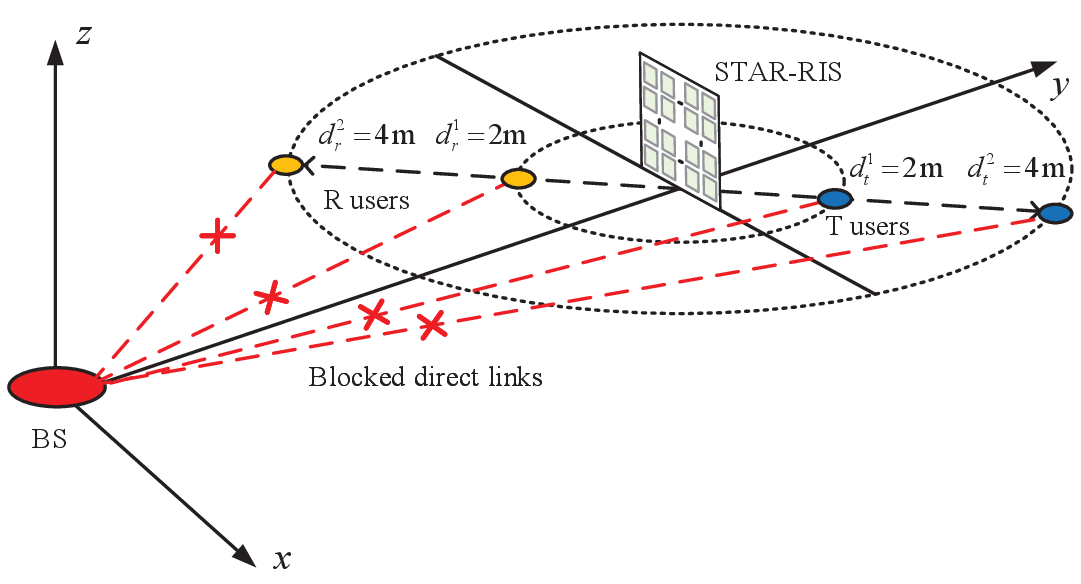}}
\caption{Two considered user setups.}\label{user_setup}\vspace{-0.2cm}
\end{figure}

We consider two user setups in simulations. 
\begin{enumerate}
\item \textbf{Random user setup:} Users in the same region line in different angles to the STAR-RIS, as shown in Fig.~\ref{Random_user}.
\item \textbf{Inline user setup:} All users in the same region line in the same angle to the STAR-RIS, as shown in Fig.~\ref{Inline_user}.
\end{enumerate}

\subsection{Convergence Behavior of BCD Algorithm}
\begin{figure}[!htbp]\vspace{-0.4cm}
    \begin{center}
        \includegraphics[width=2.7in]{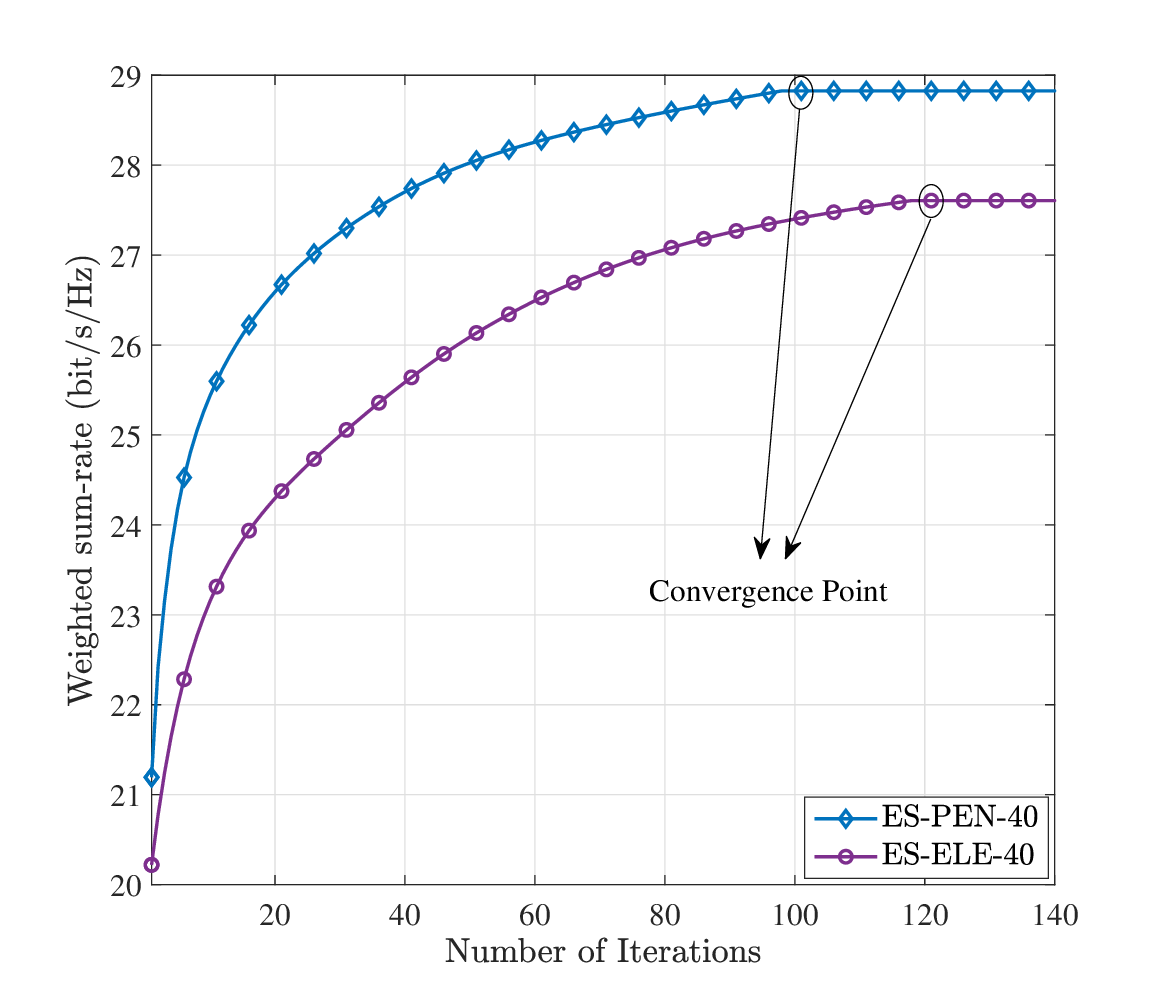}\vspace{-0.2cm}
        \caption{Convergence behavior of the proposed algorithms in near-field STAR-RIS aided multi-user communication systems.}
        \vspace{-0.5cm}
        \label{Convergence}
    \end{center}
\end{figure}

We ﬁrst study the convergence behaviour of the proposed BCD-based algorithms. In Fig.~\ref{Convergence}, we show the convergence speed and the achievable weighted sum rate of these algorithms when $N=40$. We can observe from the figure that the BCD-ELE algorithm takes more iterations to converge than the BCD-PEN algorithm. The reason behind this is the BCD-PEN algorithm updates the configuration of the STAR-RIS as a whole in each iteration while the BCD-ELE algorithm updates the configuration of the STAR-RIS in an element-wise manner in each iteration. Therefore, the BCD-ELE algorithm in general requires more iterations to converge than the BCD-PEN algorithm. Nevertheless, the BCD-ELE algorithm is still more efficient than the BCD-PEN algorithm since the PEN algorithm used in every iteration of the BCD-PEN algorithm has a significantly larger computation complexity than the ELE algorithm used in every iteration of the BCD-ELE algorithm. 
\subsection{Weighted Sum Rate of Users Versus the Transmit Power}
\begin{figure}[!htbp]\vspace{-0.5cm}
    \begin{center}
        \includegraphics[width=2.7in]{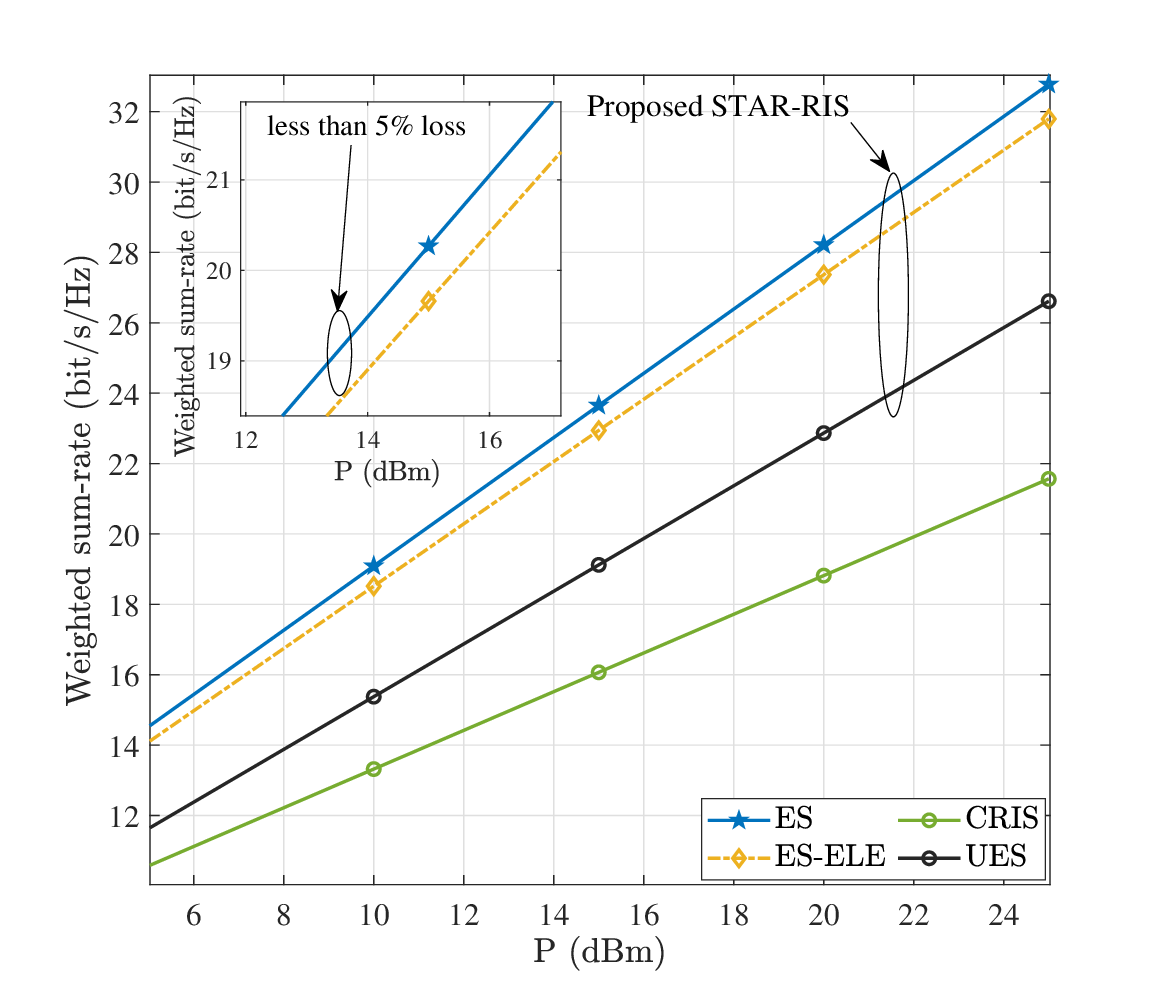}\vspace{-0.3cm}
        \caption{Weighted Sum Rate of Users Versus the Transmit Power.}\vspace{-0.5cm}
        \label{PEN_ELE_UES_CRIS}
    \end{center}
\end{figure}
In Fig.~\ref{PEN_ELE_UES_CRIS}, we investigate the weighted sum rate of users versus the transmit power of the BS station. We set $N=40$ and adopt the random user setup in the simulation. As we can see from the figure, the weighted sum rate of user for all schemes and protocols increases as the BS transmit power increases. This is expected since more power budget at the BS allow the users receive stronger signals. It can be observed that regardless the adopted operating protocol, the STAR-RIS always outperform conventional RIS baseline. Besides, Fig.~\ref{PEN_ELE_UES_CRIS} shows that ES is preferable in the investigated system. Moreover, Fig.~\ref{PEN_ELE_UES_CRIS} also reveals that the low complexity BCD-ELE algorithm achieves almost the same performance as the BCD-PEN algorithm. However, compared with BCD-PEN algorithm, the computation complexity of the BCD-ELE algorithm increases linearly with the number of STAR elements $N$, which makes it more appealing in near-field communication where the STAR array is large.
\begin{figure}[!htbp]\vspace{-0.4cm}
    \begin{center}
        \includegraphics[width=2.7in]{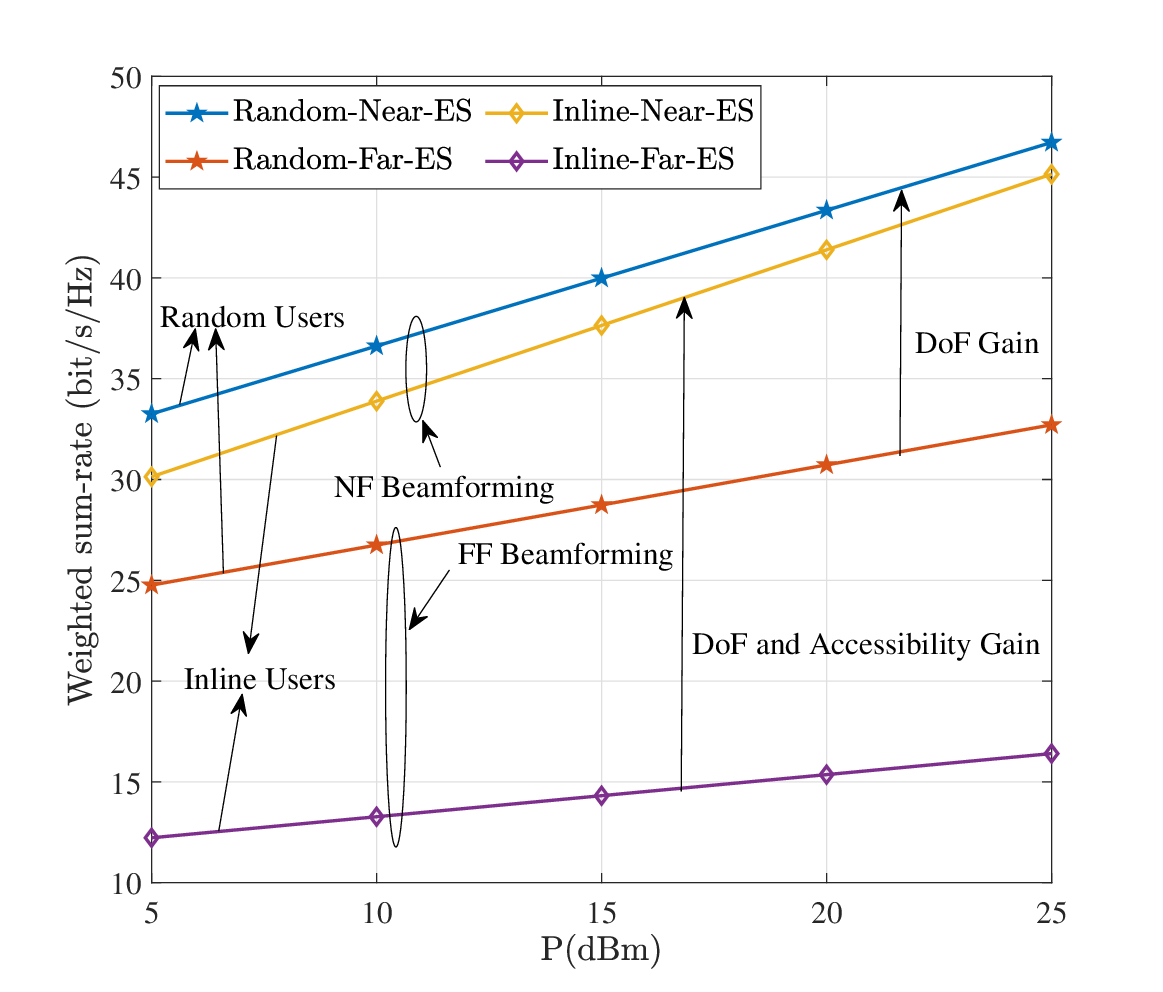}\vspace{-0.3cm}
        \caption{Weighted Sum Rate of Users Versus the Transmit Power under Different User Setups with $N=400$.}\vspace{-0.5cm}
        \label{400_inline_random}
    \end{center}
\end{figure}\\
\indent As can be seen from Fig.~\ref{400_inline_random}, for the same beamforming scheme (far-field beamforming or near-field beamforming), the random user setup always outperforms the inline user setup. This is expected since the inline user setup will lead to high inter-user interference which inevitably leads to sum rate degradation. For the same user setup, it can be observed from Fig.\ref{400_inline_random} that the near-field beamforming always provides higher capacity than the far-field beamforming. Specifically, under the random user setup, the capacity gain mainly comes from the DoFs enhancement of near-field channel, \ie, the near-field \los channels have higher rank and can bear more data streams to the multi-antenna users. Under the inline user setup, the capacity gain comes from both the DoFs enhancement and the accessibility improvement of the near-field channel, \ie, except for the gain brought by the high rank near-field \los channels, the user distance information carried by the near-field channel is leveraged by the joint beamforming algorithm to mitigate the severe inter-user interference in this case. 
\section{Conclusions}
Aiming at maximizing the weighted sum rate for the STAR-RIS aided near-field MIMO communication system, we presented BCD-based algorithms for joint optimization of beamforming matrices at the BS and TRCs at the STAR-RIS. Our simulation results demonstrated that near-field beamforming could significantly enhance the weighted sum rate for STAR-RIS aided multi-user MIMO systems.

\section{Acknowledgments}
This work is partially supported by the National Key Research and Development Project under Grant 2020YFB1806805 and Science and Technology on Communication Networks Laboratory. The work of Li Haochen was supported by China Scholarship Council.

\bibliographystyle{IEEEtran}
\bibliography{myref}

\end{document}